**Searching for Synergism Among Combinations of Drugs of Abuse and the Use of Isobolographic Analysis**

Ronald J. Tallarida[1], Uros Midic[2], Neil S. Lamarre[1], and Zoran Obradovic[2]

[1]Department of Pharmacology and Center for Substance Abuse Research

Temple University School of Medicine

3420 N. Broad Street, Philadelphia, PA 19140

[2]Department of Computer and Information Sciences and Center for Data Analytics and

Biomedical Informatics

Temple University, 303 Wachman Hall, Philadelphia, PA 19122

**Corresponding author:**

Ronald J. Tallarida, Ph.D.
ronald.tallarida@temple.edu
Tel. 215-707-3243
Fax 215-707-7068

**Abstract**

It is well known that individuals who abuse drugs usually use more than one substance. Toxic consequences of single and multiple drug use are well documented in the Treatment Episodes Data Set (TEDS) that lists combinations that result in hospital admissions. Using this list as a guide, we focused our attention on combinations that result in the most hospital admissions and searched the PubMed database to determine the number of publications dealing with these toxic combinations. Of special interest were those publications that looked for or used the term *synergism* in their titles or abstracts, a search that produced an extensive list of published articles. However, a further intersection of these with the term *isobole* revealed a surprisingly small number of literature reports. Because the method of isoboles is the most common quantitative method for distinguishing between drug synergism and simple additivity, the small number of investigations that actually employed this quantitation suggests that the term *synergism* is not properly documented in describing the toxicity among these abused substances. The possible reasons for this lack of quantitation may be related to a misunderstanding of the modeling equations. The theory and modeling are discussed here.

**1. Introduction**

It is well known that alcohol, cocaine, opioids, marijuana, and various stimulants are prominent among substances that are frequently abused. These substances have been extensively studied and the results of that effort are represented in a vast body of publications. It is also well known that drug abusers do not usually confine their usage to a single drug. In that regard, information from the Treatment Episode Data Set (TEDS) is revealing in that it provides drug combination data that resulted in hospital admissions due to drug abuse toxicity (Table 1). The TEDS data are useful and emphasize the need for even more information on drug-drug interactions among the classes of abused drugs; of specific importance are those reactions that are synergistic. Therefore, it is of interest to ask whether drug synergism has been rigorously determined for pairs of abused substances. Addressing that question is the main aim of this communication which also summarizes the theory that answers the question. Our use of the TEDS database guided us in the selection of drug combinations of interest for our further data mining of substances that are of special interest. Specifically, we examined the PubMed database to locate and count publications that include synergistic interactions for these widely abused drug combinations.

### 1.1. Synergism and Isoboles: Key words in a search strategy

A search strategy involves the use of key words and expressions that have special relevance to the search. In that regard, we included the term *synergism* as a key word that describes supra-additive interactions between two drugs. Specifically, synergism refers to effects of the drug combination (toxic or beneficial) that are numerically greater than the combination effect that is suggested by the individual drug's potency/efficacy profile. This profile is determined from each drug's dose-effect relation, where the "effect" is meant to be some common effect (therapeutic or toxic) that is produced by each. The most common method for assessing synergism is the *isobolographic* method introduced and popularized by Loewe [1-3]. In this method one identifies a particular effect of interest that is common to each drug. We then obtain the dose-effect curve of each drug and from these we derive a third graph in Cartesian coordinates that plots dose pairs that are expected to yield the specified effect at some magnitude (often 50% of the maximum effect). That third curve is the additive isobole. Because each drug individually produces the specified effect, it is reasoned that the presence of one drug reduces the dose needed of the second drug in the production of the specified effect. Therefore the graph of the dose of "drug B" against the dose of "drug A" is a monotone decreasing curve that may be linear or nonlinear. This curve is termed the additive *isobole* (See Fig. 1). All points on the curve (drug dose pairs) are expected to produce the specified effect level and, therefore, experimental points found to be below the curve mean that lesser quantities are needed to produce the effect, thereby indicating a synergistic interaction. Equivalently, synergism is indicated when a point (dose pair) on the isobole gives an effect that is greater than the specified effect for that isobole. A point on the graph that plots above the isobole indicates a sub-additive interaction. This graphical approach, which is highly dependent on the shape of the individual dose-effect curves, has been extensively examined and expanded. Since the appearance of Loewe's early works (*op.cit.*) there have appeared many theoretical and procedural details that are contained in more recent works and reviews [4-11]. The aim in all of these is the distinction between synergism and simple additivity or sub-additivity. The terms *synergism* and *isobole* are therefore key words in our data mining procedure that is described below.

## 2. Methods

Our search utilized 38 group substance names (GSNs) as listed in Table 2. Some substance names included a wild card symbol to facilitate matching to various lexical forms of the same name. Not all GSNs are disjoint; i.e., some were constructed as the unions of several other

GSNs for which the PubMed query results were small.  We initially performed searches for each of the 38 GSNs.  For each of these a query was designed to return the PubMed identifier among all entries in that database that contain at least one of the substance names in either the title or abstract.  Thus, our first search strategy counted all papers that include the substance name in the title or abstract.  This strategy assumes that if one or more of the substances are a focus of the study, then their names will at least be mentioned in the title or abstract.  We proceeded by taking into account the co-occurrences of substances, e.g., in GSN #1 we used "buprenorphine **or** "buprenex."  A further strategy counted papers that include pairs of substances from the different groups, while an additional strategy counted papers with pairs of substance names (from different groups) **and** either the term  "isobole" **or** "synergy" **or** "drug synergism."  In order to further filter the search process we searched for pairs (from different groups) **and** "isobole or "synergy."  The most refined final search included pairs of substance names **and** the term "isobole."

3.  Results

We see from Table 1 that combinations of alcohol with marijuana and alcohol with cocaine account for the greatest percentage of hospital admissions.  Also notable are combinations of cocaine and marijuana, opioids and cocaine, and various CNS stimulants with alcohol.  This group of five pairs therefore became the main targets of our further exploration in regard to reports of synergism.  (It is interesting to note that tranquilizers and other sedative hypnotics are not high in number.)  Table 2 shows the 38 search terms that were used in a list that was derived from broad Drug Enforcement Administration categories consisting of *narcotics, CNS*

*depressants, stimulants, hallucinogens, cannabis, anabolic steroids, inhalants* and *alcohol*. A search of the PubMed database revealed that there are 447,074 published papers (as of February, 2011) that include in their title or abstracts one or more names represented in this list of 38. Among this large number of reports the greatest number of publications were for category #38 alcohol (194,241), #30 testosterone (59,429), #17 benzodiazepines (41,120), #8 morphine (36,866), #18 cocaine (25,397) and #29 marijuana (22,094). Among the total of 447,074 papers there were 77,385 that included the term *synergy* or *synergism* in the title or abstract, and 1265 papers that included the term *isobole*. When the search focused on publications that contain <u>at least two substance names</u> (in title or abstract) we found 59,957 publications. Of these there were only 481 that contained either or both the terms i*sobole* or *synergy* (*synergism*). When confined to just the single term *isobole*, along with at least two substances, the number of publications dropped to 59.

When the list of 38 is viewed against the five drug combinations that resulted in the most hospital admissions (shown in Table 1) we get the numbers of publications [12-16] shown in Table 3 that deal with these toxic combinations of interest. Here we see a rather large number of publications involving alcohol with the several groups indicated. Yet, among all of these, our search showed that the term *isobole* or *isobolographic analysis* is mentioned in only three publications with these combinations, specifically each with cocaine. A somewhat similar result was found among the cocaine/marijuana publications. Here there was only one paper, and there was only one paper dealing with cocaine and opioids. These few are referenced in the table.

**4. Discussion**

Our data mining effort was extensive and revealed a very large number of publications, both clinical and preclinical, that deal with the major drugs of abuse. In our total database more than 77,000 of these publications made reference to synergism. This is a term that is very frequently associated with the toxicity of drug combinations. There is no doubt that drug combinations can be dangerous, whether the interaction is synergistic or simply additive. Yet it is important to look more closely into the use of the term synergism, a word with a very specific quantitative meaning. This kind of interaction should be used only if the drug combination has been subjected to a quantitative analysis that distinguishes between the observed combination effect and the effect that is expected from the individual drug potencies. In most cases, and certainly among the drug groups mentioned here, an analysis of synergism almost always uses isobolographic methods. Other methods of analysis such as response surface analysis [17, 5] have been used but are less common. In that regard our data mining effort shows a drastic drop to only 59 publications that include the word *isobole* in the publication title or abstract. Of course, it is possible that quantitation may have accompanied some of the papers that concluded synergism but just did not use the term isobole. Therefore no firm conclusion can be drawn. Yet the omission in the abstract of the method (isobole or other) that led to a conclusion of synergism seems unlikely if there really was a quantitative method used in the analysis. This point is further reinforced by the magnitude of the drop: over 77,000 papers mentioning synergism, but only 59 publications mentioning isobole. Drug combinations can be very useful in therapy and drug combinations can be quite important in the production of toxic reactions. Synergistic interactions are especially important in these cases of toxic reactions and also because this finding is often a first step in understanding mechanism, a fact well illustrated in the review by

Tallarida and Raffa [11] that describes the basis of the various methods used in quantitating drug-drug interactions.

The rather modest use of the term isobole (or employment of other quantitative methods) among the authors who use the term synergism may be due to some confusion that has surrounded the isobole method of analysis. The source of this confusion may be due to the different views of its originator, Loewe, and a subsequent analysis by Berenbaum [4] that are summarized in the latter's extensive review. Loewe was clear in suggesting that the additive isobole could be nonlinear but he justified his view with a rather loose mathematical treatment that had cumbersome notation. Berenbaum rejected Loewe's reservations by providing an argument that included a sham demonstration that is described subsequently. In short, Berenbaum took the linear equation $x/A + y/B = 1$ to be the **definition** of an additive interaction for a dose pair $(x,y)$ with individual potencies $A$ and $B$, respectively. For example, if the specified effect level is 50% of the maximum, then $A = ED_{50}$ of drug A and $B = ED_{50}$ of drug B. However, this is NOT the definition; it is, instead, a consequence of the fact that two agonist drugs have a constant relative potency as we now show. In other words, dose $x$ for drug A and dose $y$ for drug B give a ratio $R = A/B$ (ratio of $ED_{50}$'s) that is the same for all equally effective $x,y$ pairs. In this case any tested dose $x$ of drug A has a dose B-equivalent that is $x/R$. From this it follows that the dose of drug B alone that gave the specified effect (that dose denoted $B$) could be achieved by adding the actual $y$ and the equivalent, i.e., $y + x/R = B$. On re-arrangement this becomes

$$x/A + y/B = 1. \qquad (1)$$

If, however, the potency ratio $R$ is not a constant, then equation (1) does NOT hold and the additive isobole will be generally nonlinear. It is instructive to see how Berenbaum came to the conclusion that this linear relation *defines* the isobole. His approach used a sham combination, i.e., only one agent and a dilution of that agent. Those he considered as the two "drugs." From this he shows mathematically that the additive isobole is linear. But that situation does not constitute a proof because the diluted drug and the actual drug will necessarily have a constant potency ratio, a situation in which the isobole is clearly linear as shown above. If, however, the shapes of the individual dose-effect curves differ, (e.g., if the curves have different $E_{max}$ values) then the potency ratio is not constant, and it is easy to show that the isobole is not linear [8-11]. Hence, the Berenbaum assertion did not include the more general case of nonparallel dose effect curves, and therefore Loewe's original assertion that isoboles can be nonlinear is correct. The fact that isoboles can be (and often are) nonlinear should not detract from their usage in quantitating drug combinations since the nonlinear isobole represents no major mathematical challenge. An interaction model or method that starts with the linear form above as the *definition* of additivity will be very restricted in its use. Further, the confusion resulting from Berenbaum's rejection of Loewe's general case might explain, in part, the relatively small use of isoboles included among the many publications detected in our search.

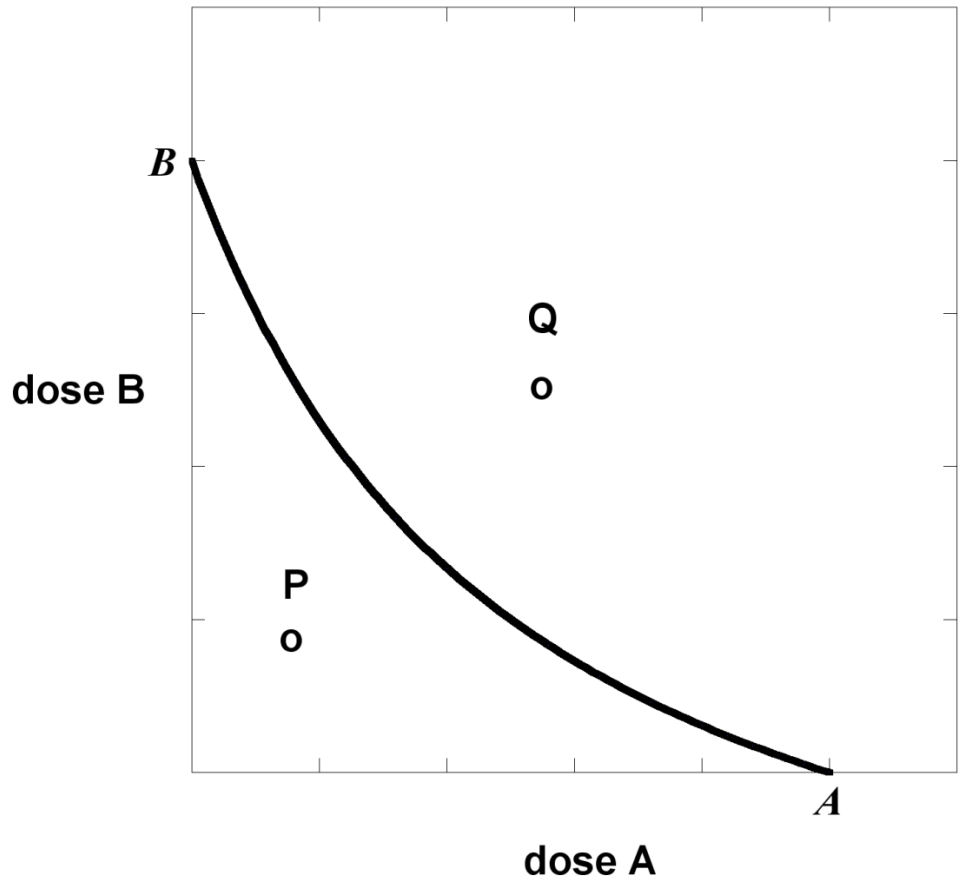

**Fig. 1.** The smooth curve is an isobole, a monotone decreasing curve of dose pairs of drugs A and B that individually produce the same effect and which are used in combination in order to produce a specified effect level (usually 50% of the maximum). The curve shows that the presence of the second drug reduces the needed dose of the first when both are present and there is no interaction between the two drugs, a situation termed "additive." The intercepts $A$ and $B$ denote the individual doses needed to achieve the effect. The shape of the isobole (curvature) depends on the potency ratio of the two substances. Regardless of its shape the isobole allows an assessment of synergism if the actual dose combination gives the effect with lesser quantities, such as point P, which falls below the curve. In contrast, an actual dose pair above the isobole, such as point Q, indicates a sub-additive interaction.

## Table 1. Treatment Episode Data Set (TEDS)
Below are substance abuse combinations by selected primary substance of abuse: TEDS number and percent distribution

| Primary substance | Secondary and tertiary substances | Number | Percent of all admissions |
|---|---|---|---|
| **All admissions** | | 1,882,584 | 100.0 |
| **Alcohol** | | 807,939 | 42.9 |
| *No other substance* | n/a | 444,781 | 23.6 |
| *1 other substance* | Marijuana | 117,272 | 6.2 |
| | Cocaine | 87,674 | 4.7 |
| | Opiates | 17,556 | 0.9 |
| | Stimulants | 12,256 | 0.7 |
| | Other | 12,027 | 0.6 |
| *2 other substances* | Cocaine & Marijuana | 53,499 | 2.8 |
| | Marijuana & Stimulants | 16,047 | 0.9 |
| | Cocaine & Opiates | 16,777 | 0.9 |
| | Marijuana & Opiates | 6,285 | 0.3 |
| | Cocaine & Stimulants | 4,400 | 0.2 |
| | Stimulants & Opiates | 1,222 | 0.1 |
| | Marijuana & Other | 9,460 | 0.5 |
| | Cocaine & Other | 4,755 | 0.3 |
| | Opiates & Other | 2,875 | 0.2 |
| | Stimulants & Other | 1,053 | 0.1 |
| **Cocaine** | | 241,699 | 12.8 |
| *No other substance* | n/a | 71,123 | 3.8 |
| *1 other substance* | Alcohol | 70,520 | 3.7 |
| | Marijuana | 23,074 | 1.2 |
| | Opiates | 6,502 | 0.3 |
| | Stimulants | 1,980 | 0.1 |
| | Other | 2,362 | 0.1 |
| *2 other substances* | Alcohol & Marijuana | 44,874 | 2.4 |
| | Opiates & Alcohol | 6,774 | 0.4 |
| | Stimulants & Alcohol | 2,587 | 0.1 |
| | Opiates & Marijuana | 2,748 | 0.1 |
| | Stimulants & Marijuana | 2,065 | 0.1 |
| | Opiates & Stimulants | 498 | * |
| | Alcohol & Other | 3,349 | 0.2 |
| | Marijuana & Other | 2,116 | 0.1 |
| | Opiates & Other | 844 | * |
| | Stimulants & Other | 283 | * |
| **Opiates** | | 331,272 | 17.6 |
| *No other substance* | n/a | 141,565 | 7.5 |
| *1 other substance* | Cocaine | 54,426 | 2.9 |
| | Alcohol | 33,576 | 1.8 |
| | Marijuana | 14,277 | 0.8 |
| | Stimulants | 3,370 | 0.2 |
| | Other | 10,950 | 0.6 |
| *2 other substances* | Cocaine & Alcohol | 30,630 | 1.6 |
| | Cocaine & Marijuana | 12,733 | 0.7 |
| | Alcohol & Marijuana | 10,741 | 0.6 |
| | Cocaine & Stimulants | 2,871 | 0.2 |
| | Stimulants & Alcohol | 1,743 | 0.1 |
| | Stimulants & Marijuana | 1,265 | 0.1 |
| | Cocaine & Other | 5,145 | 0.3 |
| | Alcohol & Other | 4,777 | 0.3 |
| | Marijuana & Other | 2,657 | 0.1 |
| | Stimulants & Other | 546 | * |
| **Marijuana** | | 283,527 | 15.1 |
| *No other substance* | n/a | 99,870 | 5.3 |
| *1 other substance* | Alcohol | 99,531 | 5.3 |
| | Cocaine | 11,046 | 0.6 |
| | Stimulants | 9,959 | 0.5 |
| | Opiates | 2,280 | 0.1 |
| | Other | 5,752 | 0.3 |
| *2 other substances* | Alcohol & Cocaine | 18,817 | 1.0 |
| | Alcohol & Stimulants | 13,561 | 0.7 |
| | Stimulants & Cocaine | 2,339 | 0.1 |
| | Alcohol & Opiates | 3,169 | 0.2 |
| | Cocaine & Opiates | 1,386 | 0.1 |
| | Stimulants & Opiates | 523 | * |
| | Alcohol & Other | 11,022 | 0.6 |
| | Cocaine & Other | 1,926 | 0.1 |
| | Stimulants & Other | 1,445 | 0.1 |
| | Opiates & Other | 901 | * |
| **Stimulants** | | 126,063 | 6.7 |
| *No other substance* | n/a | 37,773 | 2.0 |
| *1 other substance* | Alcohol | 19,137 | 1.0 |
| | Marijuana | 21,707 | 1.2 |
| | Cocaine | 3,052 | 0.2 |
| | Opiates | 1,396 | 0.1 |
| | Other | 1,625 | 0.1 |
| *2 other substances* | Marijuana & Alcohol | 25,135 | 1.3 |
| | Cocaine & Alcohol | 4,233 | 0.2 |
| | Cocaine & Marijuana | 4,314 | 0.2 |
| | Opiates & Alcohol | 1,280 | 0.1 |
| | Marijuana & Opiates | 1,333 | 0.1 |
| | Cocaine & Opiates | 763 | * |
| | Marijuana & Other | 2,032 | 0.1 |
| | Alcohol & Other | 1,444 | 0.1 |
| | Cocaine & Other | 519 | * |
| | Opiates & Other | 320 | * |
| **Other** | | 92,084 | 4.9 |

**SOURCE:** The Treatment Episode Data Set (TEDS) is maintained by the Office of Applied Studies, Substance Abuse and Mental Health Services Administration (SAMHSA). The TEDS system includes records for some 1.5 million substance abuse treatment admissions annually.

## TABLE 2. DRUGS OF ABUSE
Source: U.S. Department of Justice Drug Enforcement Administration

|    | Term(s) | PubMed search count* (Title and/or Abstract, unless stated otherwise) |
|----|---------|---------------------------------------------------------------------|
| 1  | buprenorphine, buprenex | 3,210 |
| 2  | butorphanol, stadol | 1,029 |
| 3  | codeine | 3,409 |
| 4  | fentanyl, duragesic | 12,089 |
| 5  | hydrocodone, vicodin | 397 |
| 6  | hydromorphone, dilaudid | 801 |
| 7  | methadone, dolophine | 8,715 |
| 8  | morphine, astramorph | 36,866 |
| 9  | oxycodone, oxycontin | 1,154 |
| 10 | propoxyphene, darvon | 828 |
| 11 | mephobarbital, mebaral | 81 |
| 12 | pentobarbital, nembutal | 13,000 |
| 13 | diazepam, valium | 16,838 |
| 14 | chlordiazepoxide, librium | 2,919 |
| 15 | alprazolam, xanax | 1,752 |
| 16 | lorazepam, ativan, temesta | 2,712 |
| 17 | benzodiazepine*, diazepam, valium, chlordiazepoxide, librium, alprazolam, xanax, lorazepam, ativan, temesta | 41,120 |
| 18 | cocaine | 25,397 |
| 19 | phentermine | 407 |
| 20 | diethylpropion | 194 |
| 21 | methamphetamine, dextroamphetamine | 6,579 |
| 22 | methylphenidate, dexmethylphenidate, ritalin, adderall | 4,300 |
| 23 | caffeine | 19,365 |
| 24 | ketamine | 10,380 |
| 25 | mdma, methylenedioxyamphetamine | 2,709 |
| 26 | lsd, lysergic acid | 4,137 |
| 27 | mescalin, trimethoxyphenethylamine, phenethylamine | 590 |
| 28 | psilocybin | 321 |
| 29 | tetrahydrocannabinol, thc, cannab*, marijuana | 22,094 |
| 30 | testosterone, dihydrotestosterone | 59,429 |
| 31 | nitrous oxide | 11,539 |
| 32 | alkyl nitrites | 46 |
| 33 | amyl nitrite | 428 |
| 34 | butyl nitrite | 60 |
| 35 | isopropyl nitrite | 9 |
| 36 | isobutyl nitrite | 79 |
| 37 | inhalant*, nitrous oxide, alkyl nitrites, amyl nitrite, butyl nitrite, isopropyl nitrite, isobutyl nitrite | 14,969 |
| 38 | alcohol | 194,241 |

*In the above the column total is 524,193.  When corrected for duplicates the sum is 447,074 as reported in the text.

**Table 3. Number of publications of the named drug and combinations using isoboles**

| Combination | Relevant Publications | Isobolar analysis |
|---|---|---|
| <u>Alcohol</u> with marijuana (#38 with #29) | 4326 out of 11589 (on alcohol & other drugs) | 0 |
| <u>Alcohol</u> with cocaine (#38 with #18) | 3245 out of 11589 (on alcohol and other drugs) | 3[12-14] |
| <u>Alcohol</u> + other stimulants (#38 with #19-25) | 1691 out of 11589 (on alcohol and other drugs) | 0 |
| <u>Cocaine</u> with marijuana (#18 with #29) | 2435 out of 7516 (cocaine and other drugs) | 1[15] |
| <u>Cocaine</u> with opioids (#18 with #3-9) | 2015 out of 7516 (cocaine and other drugs) | 1[16] |